\documentclass{article}

\usepackage{amsmath,amssymb,amsthm}
\usepackage{enumerate}
\newtheorem{theorem}{Theorem}

\newtheorem{definition}[theorem]{Definition}
\newtheorem{lemma}[theorem]{Lemma}
\newtheorem{proposition}[theorem]{Proposition}

\begin{document}

\title{Negations and Meets in Topos Quantum Theory\thanks{Email: \texttt{kitajima.yuichirou@nihon-u.ac.jp}
}
}


\author{Yuichiro Kitajima         
}


\maketitle

\begin{abstract}
The daseinisation is a mapping from an orthomodular lattice in ordinary quantum theory into a Heyting algebra in topos quantum theory. While distributivity does not always hold in orthomodular lattices, it does in Heyting algebras. We investigate the conditions under which negations and meets are preserved by daseinisation, and the condition that any element in the Heyting algebra transformed through daseinisation corresponds to an element in the original orthomodular lattice. We show that these conditions are equivalent, and that, not only in the case of non-distributive orthomodular lattices but also in the case of Boolean algebras containing more than four elements, the Heyting algebra transformed from the orthomodular lattice through daseinisation will contain an element that does not correspond to any element of the original orthomodular lattice.

\end{abstract}

\section{Introduction}
\label{section-introduction}
Birkoff and von Neumann \cite{birkhoff1936logic} interpreted mathematical operations of projections as providing logic for quantum systems. It is well known that it is an orthomodular lattice \cite{redei1998quantum}. In orthomodular lattices as well as Boolean algebras, joins $\vee$, meets $\wedge$, and negations $\perp$ are defined. Since $x \vee x^{\perp} = 1$ for any element $x$ in an orthomodular lattice, the law of excluded middle holds in an orthomodular lattice as well as in a Boolean algebra. In this sense, orthomodular lattices are similar to Boolean algebras. However, unlike Boolean algebras, distributivity does not necessarily hold in orthomodular lattices. This raises the following interpretive issues known as the `quantum breakfast'.

\begin{quotation}
[I]f you go to the quantum hotel and they offer you eggs and (bacon or sausage), you cannot expect to get (eggs and bacon) or (eggs and sausage) due to nondistributivity of `and' and `or'. As a formula, $e \wedge (b \vee s) \neq (e \wedge b) \vee (e \wedge s)$ in general in an orthomodular lattice. \cite[p. 57]{cannon2015generalisation}
\end{quotation}

If we consider this problem to be serious, the orthomodular lattice should be replaced by another algebra such that distributivity holds.

Isham and Butterfield discussed quantum theory in terms of topos theory \cite{isham1998topos,butterfield1999topos,hamilton2000topos,butterfield2002topos}. The topos approach to quantum theory has been further developed \cite{doring2008topos1,doring2008topos2,doring2008topos3,doring2008topos4,doring2010thing,cannon2015generalisation,doring2016topos,doring2018bridge,heunen2009topos,heunen2011bohrification,heunen2012bohrification}. The logic in this approach is based on a Heyting algebra. In this algebra, the law of excluded middle does not always hold, but distributivity does. Hence, the interpretive problem called `quantum breakfast' does not arise.

What relates an orthomodular lattice to a Heyting algebra is a mapping called daseinisation, introduced by D\"{o}ring and Isham \cite{doring2008topos2,doring2010thing}. According to them, `daseinisation ``brings-a-quantum-property-into-existence'' by hurling it into the collection of all possible classical snap-shots of the world provided by the category of context' \cite[p. 802]{doring2010thing}. Abelian von Neumann algebras \cite{doring2010thing} and complete Boolean algebras \cite{cannon2015generalisation} are considered as classical snap-shots. In the present paper, we use complete atomic Boolean algebras as classical snap-shots.

By daseinisation, ordinary quantum theory is transformed into topos quantum theory. If we consider the above interpretative problem to be serious and should avoid it, then transforming the orthomodular lattice into the Heyting algebra through daseinisation may be an option. Although the purpose of daseinisation is to transform non-distributive orthomodular lattices into Heyting algebra, it can also be applied to Boolean algebra which is distributive. In the present paper, we consider daseinisation in a general framework that takes both applications into account.

In Heyting algebras, distributivity holds, but the law of excluded middle does not necessarily hold. On the other hand, in orthomodular lattices, the law of excluded middle holds, but distributivity does not necessarily hold. Since joins are preserved by daseinisation, it is inferred that daseinisation does not always preserve negations and meets. One of the objectives of the present paper is to clarify under what conditions daseinisation does not preserve negations and meets.

As Eva \cite[Section 4.2]{eva2017topos} pointed out, when an orthomodular lattice is transformed into a Heyting algebra by daseinisation, the Heyting algebra may contain a new element that does not correspond to any element of the original orthomodular lattice. Although the information of the original orthomodular lattice is not lost by daseinisation, it may be necessary to introduce new elements in order to turn it into a Heyting algebra. Eva \cite{eva2017topos} states the following.

\begin{quotation}
Even if intuitionistic logic does turn out to be philosophically preferable to orthomodular quantum logic, it looks like this advantage would have been bought at the cost of introducing a whole new class of phantom propositions that have no natural physical interpretation. Until such an interpretation has been provided, the purported philosophical benefits of the intuitionistic logic of TQT [Topos Quantum Theory] look to have been bought at an unreasonably high cost: the introduction of a class of physical propositions with no actual physical significance. \cite[p. 1174]{eva2017topos}
\end{quotation}

The purpose of the present paper is not to solve this problem, but to clarify under what conditions this problem arises. Furthermore, we aim to relate the condition of introducing a new element into the Heyting algebra that does not correspond to any element of the original orthomodular lattice to the condition that daseinisation does not preserve negations and meets. 

The structure of this paper is as follows. In Section \ref{section-preliminaries}, we give the definitions and properties of orthomodular lattices needed in the present paper. While Cannon and D\"{o}ring \cite{cannon2015generalisation} used complete Boolean subalgebras in complete orthomodular lattices to construct Heyting algebras, we use complete atomic Boolean subalgebras. To use these algebras, in Section \ref{section-complete-atomic-boolean}, we describe the properties of complete atomic Boolean algebras. Using these properties, we construct a Heyting algebra in Section \ref{section-spectral-presheaves}. In Section \ref{section-daseinisation} we describe daseinisation, a mapping from orthomodular lattices into Heyting algebras, and in Section \ref{section-upper-adjoint} we describe its inverse mapping. 

Based on these preparations, in Section \ref{section-negations-meets}, we investigate the conditions under which negations and meets are preserved by daseinisation, and the condition that any element in the Heyting algebra transformed  through daseinisation corresponds to an element in the original orthomodular lattice. We show that these conditions are equivalent, and that, not only in the case of non-distributive orthomodular lattices but also in the case of Boolean algebras containing more than four elements, the Heyting algebra transformed from the orthomodular lattice through daseinisation will contain an element that does not correspond to any element of the original orthomodular lattice. It suggests that the fact that daseinisation introduces an element that does not correspond to any element of the original orthomodular lattice is not due to a quantum property but to a property of daseinisation itself.

\section{Preliminaries}
\label{section-preliminaries}
In this section, we summarize the properties of orthomodular lattices that will be needed in this paper.

Let $\mathcal{L}$ be a lattice.  The least element and the greatest element, if they exist, are denoted by $0$ and $1$, respectively. In the present paper, we assume that any lattice has $0$ and $1$.

\begin{definition}
A lattice $\mathcal{L}$ is called orthocomplemented if there is a mapping $a \rightarrow a^{\perp}$ of $\mathcal{L}$ to $\mathcal{L}$ satisfying the following conditions:
\begin{enumerate}
\item $a \vee a^{\perp} = 1$ and $a \wedge a^{\perp}=0$,
\item $a \leq b$ implies $a^{\perp} \geq b^{\perp}$,
\item $(a^{\perp})^{\perp}=a$
\end{enumerate}
for any elements $a,b \in \mathcal{L}$.

$a \perp b$ means that $a \leq b^{\perp}$.
\end{definition}

\begin{definition}
Let $\mathcal{L}$ be an orthocomplemented lattice. If 
\begin{equation}
b=a \vee (b \wedge a^{\perp})
\end{equation}
for any elements $a,b \in \mathcal{L}$, $\mathcal{L}$ is called an orthomodular lattice.

If, for any set $\{ x_{i} | i \in I \}$, there are $\bigwedge _{i \in I} x_{i}$ and $\bigvee _{i \in I} x_{i}$, $\mathcal{L}$ is called a complete orthomodular lattice.
\end{definition}

The projections in any von Neumann algebra form a complete orthomodular lattice.

\begin{definition}
Let $\mathcal{L}$ be an orthomodular lattice, and let $a$ and $b$ be elements in $\mathcal{L}$ such that $a \leq b$.
\begin{itemize}
\item If for any $c \in \mathcal{L}$ such that $a \leq c \leq b$, $c$ is either $a$ or $b$, then we say that $b$ covers $a$.
\item If an element $a$ covers $0$, $a$ is called an atom.
\item $\mathcal{L}$ is called an atomic orthomodular lattice if, for any non-zero element $x \in \mathcal{L}$, there is an atom $a \in \mathcal{L}$ such that $a \leq x$.
\item $\mathcal{L}$ is called an atomless orthomodular lattice if, for any non-zero element $x \in \mathcal{L}$, there is an element $y \in \mathcal{L}$ such that $0 < y < x$.
\end{itemize}
\end{definition}

\begin{definition}
Let $\mathcal{L}$ be an orthomodular lattice, and let $a$, $b$, and $c$ be elements in $\mathcal{L}$. 
\begin{itemize}
\item We say $\{a, b, c \}$ is distributive if
\begin{equation}
\begin{split}
&(a \vee b) \wedge c = (a \wedge c) \vee (b \wedge c), \\
&(a \wedge b) \vee c = (a \vee c) \wedge (b \vee c).
\end{split}
\end{equation}
\item We say $a$ commutes with $b$, in symbols $aCb$, if 
\begin{equation}
a=(a \wedge b) \vee (a \wedge b^{\perp}).
\end{equation}
\end{itemize}
\end{definition}

When $\{ a, b, c \}$ is not distributive, the interpretive problem called `quantum breakfast' described in Section \ref{section-introduction} occurs.

As you can see from the following Proposition, distributivity is related to commutativity.

\begin{proposition} \cite[p. 25]{kalmbach1983orthomodular}
\label{distributive-condition}

Let $\mathcal{L}$ be an orthomodular lattice, and let $a$, $b$, and $c$ be elements in $\mathcal{L}$ such that $bCa$ and $cCa$. Then $\{a, b, c \}$ is distributive.
\end{proposition}

There are some conditions that are equivalent to commutativity.

\begin{proposition}
\label{prop-commute}
\cite[p. 23 and p. 26]{kalmbach1983orthomodular}

Let $\mathcal{L}$ be an orthomodular lattice. For any elements $a, b \in \mathcal{L}$, the following statements are equivalent:
\begin{enumerate}
\item $aCb$,
\item $bCa$,
\item $a \wedge (a^{\perp} \vee b) = a \wedge b$,
\item $(a \wedge b) \vee (a \wedge b^{\perp}) \vee (a^{\perp} \wedge b) \vee (a^{\perp} \wedge b^{\perp})=1$.
\end{enumerate}
\end{proposition}

There are two extreme cases of orthomodular lattices: Boolean algebras and irreducible orthomodular lattices.

\begin{definition}
Let $\mathcal{L}$ be an orthomodular lattice. Define
\[ C(\mathcal{L}) := \{ x \in \mathcal{L} | \{ x, a, b \} \ \text{is distributive for all} \ a, b \in \mathcal{L} \}. \]
$C(\mathcal{L})$ is called the center of $\mathcal{L}$.
\begin{itemize}
\item We say $\mathcal{L}$ is a Boolean algebra if $C(\mathcal{L}) = \mathcal{L}$.
\item We say $\mathcal{L}$ is irreducible if $C(\mathcal{L}) = \{0, 1 \}$.
\end{itemize}
\end{definition}

Let $x$ be an element in an irreducible orthomodular lattice $\mathcal{L}$. Then there are elements $a$ and $b$ in $\mathcal{L}$ such that $\{ x, a, b \}$ is not distributive. On the other hand, for any elements $x$, $y$, and $z$ in a Boolean algebra, $\{ x, y, z \}$ is distributive.

\begin{definition}

\

\begin{itemize}
\item Let $\mathcal{L}_1$ and $\mathcal{L}_2$ be an orthomodular lattice, and let $\lambda$ be a map of $\mathcal{L}_1$ to $\mathcal{L}_2$. 

$\lambda$ is called a lattice homomorphism if
\begin{equation}
\begin{split}
&\lambda(a \vee b) = \lambda(a) \vee \lambda(b),  \\
&\lambda(a \wedge b) = \lambda(a) \wedge \lambda(b)
\end{split}
\end{equation}
for any elements $a, b \in \mathcal{L}_1$.

If $\mathcal{L}_2$ is the two element lattice $\{ 0, 1 \}$, $\lambda$ is called a two-valued homomorphism of $\mathcal{L}_1$.
\item Let $\mathcal{L}_1$ and $\mathcal{L}_2$ be a complete orthomodular lattice, and let $\lambda$ be a map of $\mathcal{L}_1$ to $\mathcal{L}_2$. 

$\lambda$ is called a completely additive lattice homomorphism if
\begin{equation}
\begin{split}
&\lambda \left( \bigvee_{i \in I} x_{i} \right) = \bigvee_{i \in I} \lambda(x_{i}), \\
&\lambda \left( \bigwedge_{i \in I} x_{i} \right) = \bigwedge_{i \in I} \lambda(x_{i}) \\
\end{split}
\end{equation}
for any subset $\{ x_{i} \in \mathcal{L} | i \in I \}$ in $\mathcal{L}$.

If $\mathcal{L}_2$ is the two element lattice $\{ 0, 1 \}$, $\lambda$ is called a completely additive two-valued homomorphism of $\mathcal{L}_1$.
\end{itemize}
\end{definition}

Let $\mathcal{L}$ be an orthomodular lattice, and let $\lambda$ be a two-valued homomorphism of $\mathcal{L}$. Clearly $\lambda(a) \leq \lambda(b)$ if $a \leq b$. Thus $\lambda(1)=1$ and $\lambda(0)=0$. Since $1=\lambda(a \vee a^{\perp})=\lambda(a) \vee \lambda(a^{\perp})$ and $0=\lambda(a \wedge a^{\perp})=\lambda(a) \wedge \lambda(a^{\perp})$, $\lambda(a)^{\perp}=\lambda(a^{\perp})$.

A filter plays an important role in examining two-valued homomorphisms.

\begin{definition}
Let $\mathcal{L}$ be an orthomodular lattice. A nonempty subset $F$ of $\mathcal{L}$ is called a filter in $\mathcal{L}$ if $a \in F$ and $b \in F$ imply $a \wedge b \in F$ and $a \in F$ and $a \leq c$ imply $c \in F$. 
\end{definition}

\begin{proposition}
\label{prop-redei}
\cite[Proposition 3.9, Propositon 3.11]{redei1998quantum}

Let $\mathcal{B}$ be a complete Boolean algebra and let $\mathbb{F}$ be a filter in $\mathcal{B}$. The following are equivalent:
\begin{enumerate}
\item For every $x \in \mathcal{B}$, either $x \in \mathbb{F}$ or $x^{\perp} \in \mathbb{F}$,
\item For some two-valued homomorphism $\lambda$ of $\mathcal{B}$, $\mathbb{F}=\lambda^{-1}(1)$.
\end{enumerate}
\end{proposition}

\section{Complete atomic Boolean algebras}
\label{section-complete-atomic-boolean}

Cannon and D\"{o}ring \cite{cannon2015generalisation} constructed Heyting algebras using complete Boolean subalgebras in complete orthomodular lattices and a two-valued homomorphism of these subalgebras. This homomorphism is not necessarily completely additive \cite[Proposition 3.1 and Proposition 4.1]{kitajima2005interpretations}. For example, there are a two-valued homomorphism $\lambda$ and $\{ x_{i} | i \in I \}$ such that
\begin{equation}
\lambda \left( \bigvee_{i \in I} x_{i} \right) = 1, \ \ \ \lambda(x_{i})=0
\end{equation}
for any $i \in I$. This two-valued homomorphism $\lambda$ is not completely additive.

In the present paper, only completely additive homomorphisms are used.

\begin{proposition}
\label{prop-atom}
Let $\mathcal{B}$ be a complete atomic Boolean algebra, let $\text{ca}\Sigma_{\mathcal{B}}$ be the set of completely additive two-valued homomorphisms of $\mathcal{B}$, and let $\mathbb{A}(\mathcal{B})$ be the set of atoms in $\mathcal{B}$.

Then there is an isomorphism $\beta_{\mathcal{B}}$ from $\mathbb{A}(\mathcal{B})$ to $\text{ca}\Sigma_{\mathcal{B}}$ such that for any $a \in \mathbb{A}(\mathcal{B})$
\[ \beta_{\mathcal{B}}(a)=\lambda_{a}, \]
where
\[
\lambda_{a}(x) =
\begin{cases}
1 & (a \leq x) \\
0 & (a \wedge x = 0).
\end{cases}
\]


\end{proposition}

\begin{proof}
Let $a$ be an atom in $\mathcal{B}$, and let $\mathbb{F} := \{ x \in \mathcal{B} | a \leq x \}$. $\mathbb{F}$ is a filter.

Suppose that $x \not\in \mathbb{F}$. Then $a \wedge x = 0$ since $a$ is an atom in $\mathcal{B}$. Thus $x^{\perp} = x^{\perp} \vee (a \wedge x)=(x^{\perp} \vee a) \wedge (x^{\perp} \vee x)=x^{\perp} \vee a \geq a$, which implies $x^{\perp} \in \mathbb{F}$. 

By Proposition \ref{prop-redei}, there is a two-valued homomorphism $\lambda_{a}$ of $\mathcal{B}$ such that $\mathbb{F}=\lambda_{a}^{-1}(1)$, that is,
\begin{equation}
\label{eq-atom-0}
\lambda_{a}(x) =
\begin{cases}
1 & (a \leq x) \\
0 & (a \wedge x = 0).
\end{cases}
\end{equation}

Let $\{x_{i} \in \mathbb{F}|i \in I \}$ be a subset of $\mathbb{F}$. Then $a \leq \bigwedge_{i \in I} x_{i}$, which implies $\bigwedge_{i \in I} x_{i} \in \mathbb{F}$. Thus 
\begin{equation}
\label{eq-atom-1}
\lambda_{a} \left( \bigwedge_{i \in I} x_{i} \right) = \bigwedge_{i \in I} \lambda_{a}(x_{i}). 
\end{equation}
By Equation (\ref{eq-atom-1}),
\begin{equation}
\begin{split}
1-\lambda_{a} \left( \bigvee_{i \in I} x_{i} \right) 
&=\lambda_{a} \left(  \left( \bigvee_{i \in I}x_{i} \right)^{\perp} \right)  
=\lambda_{a} \left( \bigwedge_{i \in I} x_{i}^{\perp} \right) \\
&=\bigwedge_{i \in I} \lambda_{a}(x_{i}^{\perp}) 
=\bigwedge_{i \in I} \left(1 - \lambda_{a}(x_{i}) \right)
=1-\bigvee_{i \in I}\lambda_{a}(x_{i}).
\end{split}
\end{equation}
Thus
\begin{equation}
\label{eq-atom-2}
\lambda_{a} \left( \bigvee_{i \in I}x_{i} \right)=\bigvee_{i \in I}\lambda_{a}(x_{i}). 
\end{equation}
By Equations (\ref{eq-atom-1}) and (\ref{eq-atom-2}), $\lambda_{a}$ is completely additive.

Define
\begin{equation}
\label{eq-atom-3}
\beta_{\mathcal{B}}(a)=\lambda_{a} 
\end{equation}
for any $a \in \mathbb{A}(\mathcal{B})$. Then $\beta_{\mathcal{B}}$ is a mapping of $\mathbb{A}(\mathcal{B})$ to $\text{ca}\Sigma_{\mathcal{B}}$. 



Let $\lambda$ be an element in $\text{ca}\Sigma_{\mathcal{B}}$ and let $\mathbb{A}=\{ x \in \mathcal{B}|\lambda(x)=1 \}$. Define $a = \bigwedge_{x \in \mathbb{A}} x$. Then for any element $x$ such that $\lambda(x)=1$, $a \leq x$.
Since $\lambda$ is completely additive,
\begin{equation}
\label{eq-atom-3.9}
 1 = \bigwedge_{x \in \mathbb{A}}\lambda(x)=\lambda \left( \bigwedge_{x \in \mathbb{A}} x \right) = \lambda(a).  
\end{equation}
Thus $a \neq 0$.
Let $y$ be any element in $\mathcal{B}$ such that $0 \leq y < a$. If $\lambda(y)=1$, then $a \leq y$ since $y \in \mathbb{A}$. It is a contradiction. Thus $\lambda(y)=0$, which implies $\lambda(y^{\perp})=1$. Therefore $a \leq y^{\perp}$. It follows $y=y \wedge a \leq y \wedge y^{\perp}=0$. Therefore $a$ is an atom in $\mathcal{B}$.

For any element $x \in \mathcal{B}$ such that $a \wedge x = 0$, 
\begin{equation}
\label{eq-atom-3.99}
0=\lambda(a \wedge x)=\lambda(a) \wedge \lambda(x) = 1 \wedge \lambda(x)=\lambda(x).
\end{equation}

By Equation (\ref{eq-atom-3.9}), $\lambda(x)=1$ for any element $x \in \mathcal{B}$ such that $a \leq x$.
Thus
\begin{equation}
\label{eq-atom-4}
\lambda(x) =
\begin{cases}
1 & (a \leq x) \\
0 & (a \wedge x = 0)
\end{cases}
\end{equation}
Therefore $\beta_{\mathcal{B}}(a)=\lambda$, that is, $\beta_{\mathcal{B}}$ is surjective.

Let $a$ and $a'$ be atoms in $\mathcal{B}$ such that $\lambda_{a}=\lambda_{a'}$. Then $1=\lambda_{a'}(a')=\lambda_{a}(a')$. Thus $a \leq a'$. Since $a'$ is an atom and $a>0$, $a=a'$. Therefore $\beta_{\mathcal{B}}$ is injective.

\end{proof}

\begin{proposition}
\label{prop-element}
Let $\mathcal{B}$ be a complete atomic Boolean algebra, let $ca\Sigma_{\mathcal{B}}$ be the set of completely additive two-valued homomorphisms of $\mathcal{B}$, and let $P(ca\Sigma_{\mathcal{B}})$ be the set of all subsets of $ca\Sigma_{\mathcal{B}}$. 
\begin{enumerate}
\item A mapping $\alpha_{\mathcal{B}}: \mathcal{B} \rightarrow P(ca\Sigma_{\mathcal{B}})$ such that 
\begin{equation}
\alpha_{\mathcal{B}}(b)=\{ \lambda \in ca\Sigma_{\mathcal{B}} | \lambda(b)=1 \} 
\end{equation}
 is an isomorphism.
\item
\begin{equation}
\alpha_{\mathcal{B}} \left( \bigvee_{i \in I} x_{i} \right)=\bigcup_{i \in I}\alpha_{\mathcal{B}}(x_{i}) 
\end{equation}
for any subset $\{ x_{i} \in \mathcal{B} | i \in I \}$.
\end{enumerate}
\end{proposition}

\begin{proof}
\begin{enumerate}
\item
Let $\mathbb{S}$ be a subset of $\text{ca}\Sigma_{\mathcal{B}}$, let $\mathbb{A}(\mathcal{B})$ be the set of atoms in $\mathcal{B}$, and let $\beta_{\mathcal{B}}$ be an isomorphism from $\mathbb{A}(\mathcal{B})$ to $\text{ca}\underline{\Sigma}$ which is defined in Proposition \ref{prop-atom}. Then, for any $\lambda \in \mathbb{S}$, there is a unique atom $\beta_{\mathcal{B}}^{-1}(\lambda)$ such that 
\begin{equation}
\label{eq-element-0.9}
\lambda(x) =
\begin{cases}
1 & (\beta_{\mathcal{B}}^{-1}(\lambda) \leq x), \\
0 & (\beta_{\mathcal{B}}^{-1}(\lambda) \wedge x = 0).
\end{cases}
\end{equation}
Thus for any $\lambda' \in \mathbb{S}$, 
\begin{equation}
\lambda' \left( \bigvee_{\lambda \in \mathbb{S}} \beta_{\mathcal{B}}^{-1}(\lambda) \right)=\bigvee_{\lambda \in \mathbb{S}}\lambda'( \beta_{\mathcal{B}}^{-1}(\lambda))=1, 
\end{equation}
which implies 
\begin{equation}
\label{eq-element-1}
\mathbb{S} \subseteq \alpha_{\mathcal{B}} \left( \bigvee_{\lambda \in \mathbb{S}} \beta_{\mathcal{B}}^{-1}(\lambda) \right). 
\end{equation}

Let $\lambda' \in \alpha_{\mathcal{B}} \left( \bigvee_{\lambda \in \mathbb{S}} \beta_{\mathcal{B}}^{-1}(\lambda) \right)$. Then
\begin{equation}
1=\lambda' \left( \bigvee_{\lambda \in \mathbb{S}} \beta_{\mathcal{B}}^{-1}(\lambda) \right) = \bigvee_{\lambda \in \mathbb{S}}  \lambda'(\beta_{\mathcal{B}}^{-1}(\lambda)).
\end{equation}
Thus there is an atom $\beta_{\mathcal{B}}^{-1}(\lambda) \in \mathcal{B}$ such that $\lambda'(\beta_{\mathcal{B}}^{-1}(\lambda))=1$ and $\lambda \in \mathbb{S}$. For any $x \in \mathcal{B}$ such that $x \geq \beta_{\mathcal{B}}^{-1}(\lambda)$, 
\begin{equation}
\label{eq-element-1.1}
1=\lambda'(\beta_{\mathcal{B}}^{-1}(\lambda)) \leq \lambda'(x). 
\end{equation}
For any $x \in \mathcal{B}$ such that $\beta_{\mathcal{B}}^{-1}(\lambda) \wedge x= 0$, 
\begin{equation}
\label{eq-element-1.2}
0=\lambda'(\beta_{\mathcal{B}}^{-1}(\lambda) \wedge x)=\lambda'(\beta_{\mathcal{B}}^{-1}(\lambda)) \wedge \lambda'(x)=1 \wedge \lambda'(x) = \lambda'(x). 
\end{equation}
By Equations (\ref{eq-element-1.1}) and (\ref{eq-element-1.2}), $\beta_{\mathcal{B}}(\beta_{\mathcal{B}}^{-1}(\lambda))=\lambda'$. Thus $\lambda' = \lambda \in \mathbb{S}$. 
It means 
\begin{equation}
\label{eq-element-2}
\mathbb{S} \supseteq \alpha_{\mathcal{B}}\left (\bigvee_{\lambda \in \mathbb{S}} \beta_{\mathcal{B}}^{-1}(\lambda) \right). 
\end{equation}
By Equations (\ref{eq-element-1}) and (\ref{eq-element-2}),
\begin{equation}
\label{eq-element-3}
\mathbb{S} = \alpha_{\mathcal{B}}\left (\bigvee_{\lambda \in \mathbb{S}} \beta_{\mathcal{B}}^{-1}(\lambda) \right). 
\end{equation}
Therefore $\alpha_{\mathcal{B}}$ is surjective.


Let $\alpha_{\mathcal{B}}(b)=\alpha_{\mathcal{B}}(b')$. For any $\lambda \in \alpha_{\mathcal{B}}(b)$, there is an atom $\beta_{\mathcal{B}}^{-1}(\lambda)$ in $\mathcal{B}$ such that 
\begin{equation}
\label{equ-element-2.99}
\lambda(x) =
\begin{cases}
1 & (\beta_{\mathcal{B}}^{-1}(\lambda) \leq x) \\
0 & (\beta_{\mathcal{B}}^{-1}(\lambda) \wedge x = 0).
\end{cases}
\end{equation}
Since $\lambda(b)=1$, $\beta_{\mathcal{B}}^{-1}(\lambda) \leq b$ by Equation (\ref{equ-element-2.99}). Thus 
\begin{equation}
\label{eq-element-3}
\bigvee_{\lambda \in \alpha_{\mathcal{B}}(b)} \beta_{\mathcal{B}}^{-1}(\lambda) \leq b. 
\end{equation}
Suppose that $\bigvee_{\lambda \in \alpha_{\mathcal{B}}(b)} \beta_{\mathcal{B}}^{-1}(\lambda) \lneq b$. Then $b \wedge \left( \bigvee_{\lambda \in \alpha_{\mathcal{B}}(b)} \beta_{\mathcal{B}}^{-1}(\lambda) \right)^{\perp} \neq 0$. Since $\mathcal{B}$ is atomic, there is an atom $c$ in $\mathcal{B}$ such that 
\begin{equation}
c \leq b \wedge \left( \bigvee_{\lambda \in \alpha_{\mathcal{B}}(b)} \beta_{\mathcal{B}}^{-1}(\lambda) \right)^{\perp}. 
\end{equation}
There is a completely additive two-valued homomorphism $\beta_{\mathcal{B}}(c)$ such that 
\begin{equation}
\beta_{\mathcal{B}}(c)(x) =
\begin{cases}
1 & (c \leq x) \\
0 & (c \wedge x = 0).
\end{cases}
\end{equation}
Then 
\begin{equation}
\label{eq-element-3.01}
\beta_{\mathcal{B}}(c) \in \alpha_{\mathcal{B}}(b)
\end{equation} since $1 = \beta_{\mathcal{B}}(c)(c) \leq \beta_{\mathcal{B}}(c)(b) \leq 1$. On the other hand,
\begin{equation} 
\begin{split}
1
&=\beta_{\mathcal{B}}(c)(c) 
\leq \beta_{\mathcal{B}}(c)\left( \left( \bigvee_{\lambda \in \alpha_{\mathcal{B}}(b)} \beta_{\mathcal{B}}^{-1}(\lambda) \right)^{\perp} \right) \\
&= \beta_{\mathcal{B}}(c) \left( \bigwedge_{\lambda \in \alpha_{\mathcal{B}}(b)} \beta_{\mathcal{B}}^{-1}(\lambda)^{\perp} \right) 
= \bigwedge_{\lambda \in \alpha_{\mathcal{B}}(b)} \beta_{\mathcal{B}}(c) \left( \beta_{\mathcal{B}}^{-1}(\lambda)^{\perp} \right)
\end{split}
\end{equation}
implies 
\begin{equation}
\label{eq-element-3.02}
\beta_{\mathcal{B}}(c) \left( \beta_{\mathcal{B}}^{-1}(\lambda) \right)=0
\end{equation} for any $\lambda \in \alpha_{\mathcal{B}}(b)$. Since $\beta_{\mathcal{B}}(c) \in \alpha_{\mathcal{B}}(b)$ by Equation (\ref{eq-element-3.01}), 
\begin{equation}
0=\beta_{\mathcal{B}}(c)\left( \beta_{\mathcal{B}}^{-1} \left(\beta_{\mathcal{B}}(c) \right) \right)=\beta_{\mathcal{B}}(c)(c)=1
\end{equation}
 by Equation (\ref{eq-element-3.02}). It is a contradiction. Thus, 
\begin{equation}
\label{eq-element-5}
\bigvee_{\lambda \in \alpha_{\mathcal{B}}(b)} \beta_{\mathcal{B}}^{-1}(\lambda) = b. 
\end{equation}
Similarly, 
\begin{equation}
\label{eq-element-6}
\bigvee_{\lambda \in \alpha_{\mathcal{B}}(b')} \beta_{\mathcal{B}}^{-1}(\lambda) = b'. 
\end{equation}
By Equations (\ref{eq-element-5}) and (\ref{eq-element-6}),
\begin{equation}
 b=\bigvee_{\lambda \in \alpha_{\mathcal{B}}(b)} \beta_{\mathcal{B}}^{-1}(\lambda) =\bigvee_{\lambda \in \alpha_{\mathcal{B}}(b')} \beta_{\mathcal{B}}^{-1}(\lambda) =b'. 
\end{equation}
Therefore $\alpha_{\mathcal{B}}$ is injective.

\item
Let $\lambda \in \bigcup_{i \in I} \alpha_{\mathcal{B}}(x_{i})$. Then for some $i \in I$, $\lambda \in \alpha_{\mathcal{B}}(x_{i})$. Thus $1=\lambda(x_{i}) \leq \lambda \left( \bigvee_{i \in I} x_{i} \right) \leq 1$, which implies $\lambda \in \alpha_{\mathcal{B}} \left( \bigvee_{i \in I} x_{i} \right)$. Therefore
\begin{equation}
\label{eq-element-8}
 \alpha_{\mathcal{B}} \left( \bigvee_{i \in I} x_{i} \right) \supseteq \bigcup_{i \in I}\alpha_{\mathcal{B}}(x_{i}). 
\end{equation}
Let $\lambda \in \alpha_{\mathcal{B}} \left( \bigvee_{i \in I} x_{i} \right)$. Then $1 = \lambda \left( \bigvee_{i \in I} x_{i} \right) = \bigvee_{i \in I} \lambda \left( x_{i} \right)$ since $\lambda$ is completely additive. It means $\lambda(x_{i})=1$ for some $i \in I$. Thus $\lambda \in \alpha_{\mathcal{B}}(x_{i}) \subseteq \bigcup_{i \in I} \alpha_{\mathcal{B}}(x_{i})$. Therefore
\begin{equation}
\label{eq-element-9}
 \alpha_{\mathcal{B}} \left( \bigvee_{i \in I} x_{i} \right) \subseteq \bigcup_{i \in I}\alpha_{\mathcal{B}}(x_{i}). 
\end{equation}
By Equations (\ref{eq-element-8}) and (\ref{eq-element-9}),
\begin{equation}
\label{eq-element-9}
 \alpha_{\mathcal{B}} \left( \bigvee_{i \in I} x_{i} \right) = \bigcup_{i \in I}\alpha_{\mathcal{B}}(x_{i}). 
\end{equation}
\end{enumerate}
\end{proof}

\section{Completely additive spectral presheaves}
\label{section-spectral-presheaves}

A Heyting algebra can be constructed using complete atomic Boolean algebras. In this section, we will introduce implications and negations in this Heyting algebra.

\begin{definition}
Let $\mathcal{L}$ be a complete orthomodular lattice, and let $\mathbb{B}(\mathcal{L})$ be the set of all complete atomic Boolean subalgebras of $\mathcal{L}$.
The category $\mathbb{B}(\mathcal{L})$ has
\begin{itemize}
\item Objects: $\mathcal{B} \in \mathbb{B}(\mathcal{L})$,
\item Morphisms: For any $\mathcal{B}', \mathcal{B} \in \mathbb{B}(\mathcal{L})$, there exists an arrow between them $i_{\mathcal{B}'\mathcal{B}}: \mathcal{B}' \rightarrow \mathcal{B}$ iff $\mathcal{B}' \subseteq \mathcal{B}$.
\end{itemize}

The completely additive spectral presheaf $\text{ca}\underline{\Sigma}$ of $\mathcal{L}$ is the contravariant functor $\mathbb{B}(\mathcal{L})^{op} \rightarrow \textbf{Sets}$ defined by:
\begin{itemize}
\item Objects: Given an object $\mathcal{B}$ in $\mathbb{B}(\mathcal{L})$, $\text{ca}\underline{\Sigma}(\mathcal{B})$ is the set of all completely additive homomorphism of $\mathcal{B}$,
\item Morphisms: Given a morphism $i_{\mathcal{B}'\mathcal{B}}: \mathcal{B}' \rightarrow \mathcal{B}$ ($\mathcal{B}' \subseteq \mathcal{B})$, $\text{ca}\underline{\Sigma}(i_{\mathcal{B}'\mathcal{B}}) : \text{ca}\underline{\Sigma}(\mathcal{B}) \rightarrow \text{ca}\underline{\Sigma}(\mathcal{B}')$ is defined by
\begin{equation}
\text{ca}\underline{\Sigma}(i_{\mathcal{B}'\mathcal{B}})(\lambda) := \lambda|_{\mathcal{B}'}  
\end{equation}
for any $\lambda \in \text{ca}\underline{\Sigma}(\mathcal{B})$.
\end{itemize}
\end{definition}

\begin{definition}
Let $\mathcal{L}$ be a complete orthomodular lattice, and let $\mathbb{B}(\mathcal{L})$ be the set of complete atomic Boolean subalgebras of $\mathcal{L}$.
A subobject $\underline{S}$ of the completely additive spectral presheaf $\text{ca}\underline{\Sigma}$ is a contravariant functor $\underline{S}: \mathbb{B}(\mathcal{L}) \rightarrow \textbf{Sets}$ such that:
\begin{itemize}
\item $\underline{S}(\mathcal{B})$ is a subset of $\text{ca}\underline{\Sigma}(\mathcal{B})$ for all $\mathcal{B} \in \mathbb{B}(\mathcal{L})$,
\item Given a morphism $i_{\mathcal{B}'\mathcal{B}}$, then $\underline{S}(i_{\mathcal{B}'\mathcal{B}}): \underline{S}(\mathcal{B}) \rightarrow \underline{S}(\mathcal{B}')$ is defined by
\begin{equation}
\underline{S}(i_{\mathcal{B}'\mathcal{B}})(\lambda) := \lambda|_{\mathcal{B}'}   \end{equation}
for any $\lambda \in \underline{S}(\mathcal{B})$.

The set of all subobjects of $\text{ca}\underline{\Sigma}$ is denoted as $\text{Sub} \text{ca}\underline{\Sigma}$.
\end{itemize}
\end{definition}

$\text{Subca}\underline{\Sigma}$ is a Heyting algebra. The explicit description of the operation $\wedge$, $\vee$, $\underline{0}$, $\underline{1}$, $\Rightarrow$, and $\neg$ is as follows \cite[p. 56]{maclane2012sheaves} \cite[Theorem 15]{doring2010thing}.

\begin{definition}
We define a partial order on $\text{Subca}\underline{\Sigma}$:
\begin{equation}
\forall \underline{S}, \underline{T} \in \text{Sub} \text{ca}\underline{\Sigma}: \underline{S} \leq \underline{T} \Longleftrightarrow (\forall \mathcal{B} \in \mathbb{B}(\mathcal{L}): \underline{S}(\mathcal{B}) \subseteq \underline{T}(\mathcal{B})).
\end{equation}
Meets and joins are given by set-theoretic intersections and unions, respectively. They are defined as follows:

For any family $(\underline{S}_{i})_{i \in I}$,
\begin{equation}
\forall \mathcal{B} \in \mathbb{B}(\mathcal{L}): \left( \bigwedge_{i \in I} \underline{S}_{i} \right)(\mathcal{B}) := \bigcap_{i \in I} \underline{S}_{i}(\mathcal{B}),  
\end{equation}
\begin{equation}
\forall \mathcal{B} \in \mathbb{B}(\mathcal{L}): \left( \bigvee_{i \in I} \underline{S}_{i} \right)(\mathcal{B}) := \bigcup_{i \in I} \underline{S}_{i}(\mathcal{B}). 
\end{equation}

The unit element $\underline{1}$ is $\text{ca}\underline{\Sigma}$, and the zero element $\underline{0}$ is the subobject of $\text{ca}\underline{\Sigma}$ such that  $\underline{0}(\mathcal{B})=\emptyset$ for any $\mathcal{B} \in \mathbb{B}(\mathcal{L})$.

The Heyting implication $\underline{S} \Rightarrow \underline{T}$ is defined as follows.
\begin{equation}
(\underline{S} \Rightarrow \underline{T})(\mathcal{B}) := \{ \lambda \in \text{ca}\underline{\Sigma}(\mathcal{B}) | \forall \mathcal{B}' \subseteq \mathcal{B} \ \ \text{: if} \ \lambda|_{\mathcal{B}'} \in \underline{S}, \ \text{then} \ \lambda|_{\mathcal{B}'} \in \underline{T} \}
\end{equation}
for any $\mathcal{B} \in \mathbb{B}(\mathcal{L})$.

The Heyting negation $\neg \underline{S}$ is defined as follows.
\begin{equation}
\neg \underline{S} : = \underline{S} \Rightarrow \underline{0}.
\end{equation}
For any $\mathcal{B} \in \mathbb{B}(\mathcal{L})$,
\begin{equation}
\label{eq-heyting-negation-0.01}
\neg \underline{S}(\mathcal{B})=\{\lambda \in \text{ca}\underline{\Sigma}(\mathcal{B}) | \forall \mathcal{B}' \subseteq \mathcal{B} \ : \ \lambda|_{\mathcal{B}'} \not\in \underline{S}(\mathcal{B}') \}.
\end{equation}
\end{definition}

Cannon and D\"{o}ring \cite{doring2016topos,cannon2015generalisation} pointed out that there is a different kind of an implication than the Heyting implication $\Rightarrow$.
\begin{equation}
\begin{split}
\left( \underline{S} \vee \bigwedge_{i \in I} \underline{R_{i}} \right)(\mathcal{B})
&=\underline{S}(\mathcal{B}) \cup \left( \bigcap_{i \in I} \underline{R_{i}}(\mathcal{B}) \right)  \\
&=\bigcap_{i \in I} \left( \underline{S}(\mathcal{B})  \cup \underline{R_{i}}(\mathcal{B}) \right)  \\
&=\bigwedge_{i \in I} \left(  \underline{S}  \vee \underline{R_{i}} \right)(\mathcal{B})
\end{split}
\end{equation}
for any family $\{ \underline{R_{i}} | i \in I \} \subseteq \text{Subca}\underline{\Sigma}$ and any $\mathcal{B} \in \mathbb{B}(\mathcal{L})$. Hence, for any $\underline{S}$ the functor
\begin{equation}
\underline{S} \vee \_ : \text{Subca}\underline{\Sigma} \rightarrow \text{Subca}\underline{\Sigma}
\end{equation}
preserves all meets. By the adjoint functor theorem \cite[Theorem 3.3.3 and Example 3.3.9e]{borceux1994handbook1}, it has a left adjoint
\begin{equation}
\underline{S} \Leftarrow \_ : \text{Subca}\underline{\Sigma} \rightarrow \text{Subca}\underline{\Sigma}
\end{equation}
and
\begin{equation}
(\underline{S} \Leftarrow \underline{T}) \leq \underline{R} \ \ \text{iff} \ \ \underline{S} \leq \underline{T} \vee \underline{R}.
\end{equation}
Thus
\begin{equation}
(\underline{S} \Leftarrow \underline{T}) = \bigwedge \{ \underline{R} \in \text{Subca}\underline{\Sigma} | \underline{S} \leq \underline{T} \vee \underline{R} \}.
\end{equation}
$\Leftarrow$ is called the co-Heyting implication.

By using the co-Heyting implication, the co-Heyting negation $\sim \underline{T}$ is defined as follows.
\begin{equation}
\label{eq-co-negation-def}
\begin{split}
\sim \underline{T} &:= (\underline{\Sigma} \Leftarrow \underline{T}) \\
&=\bigwedge \{ \underline{R} \in \text{Subca}\underline{\Sigma} | \text{ca}\underline{\Sigma} \leq \underline{T} \vee \underline{R} \} \\
&=\bigwedge \{ \underline{R} \in \text{Subca}\underline{\Sigma} | \text{ca}\underline{\Sigma} = \underline{T} \vee \underline{R} \}.
\end{split}
\end{equation}
By Equation (\ref{eq-co-negation-def}),
\begin{equation}
\label{eq-co-negation-prop}
\sim \underline{T}(\mathcal{B}) = \{ \lambda \in \text{ca}\underline{\Sigma}(\mathcal{B})|\lambda \not\in \underline{T}(\mathcal{B}) \}
\end{equation}
for any $\mathcal{B} \in \mathbb{B}(\mathcal{L})$.

\section{Daseinisation}
\label{section-daseinisation}
The mapping from an orthomodular lattice to its completely additive spectral presheaf is daseinisation \cite{doring2008topos2,doring2010thing}. By daseinisation, an orthomodular lattice is transformed into a Heyting algebra. In this section, we give its definition and its properties.

\begin{definition}
Let $\mathcal{L}$ be a complete orthomodular lattice, and let $\mathbb{B}(\mathcal{L})$ be the set of complete atomic Boolean subalgebras of $\mathcal{L}$. For any element $x \in \mathcal{L}$ and $\mathcal{B} \in \mathbb{B}(\mathcal{L})$, $\delta(x)_{\mathcal{B}}$ is defined as follows.
\begin{equation}
\delta(x)_{\mathcal{B}} := \bigwedge \{ y \in \mathcal{B} | y \geq x \}. 
\end{equation}
\end{definition}

To prove Proposition \ref{prop-daseinisation}, we prepare the following lemma.

\begin{lemma}
\label{lemma-atom}
Let $\mathcal{L}$ be a complete orthomodular lattice, let $\mathbb{B}(\mathcal{L})$ be a set of complete atomic Boolean subalgebras in $\mathcal{L}$, let $\mathcal{B}$ and $\mathcal{B}'$ be elements in $\mathbb{B}(\mathcal{L})$ such that $\mathcal{B}' \subseteq \mathcal{B}$, and let $\mathbb{A}(\mathcal{B})$ be a set of atoms in $\mathcal{B}$. 
\begin{enumerate}
\item For any subset $\{ x_{i} \in \mathcal{L} | i \in I \}$ of $\mathcal{L}$ and any $\mathcal{B} \in \mathbb{B}(\mathcal{L})$, 
\begin{equation}
 \delta \left( \bigvee_{i \in I} x_{i} \right)_{\mathcal{B}} = \bigvee_{i \in I} \delta(x_{i})_{\mathcal{B}}. 
 \end{equation}
\item For any element $x \in \mathcal{L}$, 
\begin{equation}
\delta(x)_{\mathcal{B}'} = \bigvee \{\delta(a)_{\mathcal{B}'} | a \leq \delta(x)_{\mathcal{B}}, a \in \mathbb{A}(\mathcal{B}) \}.
\end{equation}
\item For any atom $a \in \mathbb{A}(\mathcal{B})$, $\delta(a)_{\mathcal{B}'}$ is an atom in $\mathcal{B}'$.
\end{enumerate}
\end{lemma}

\begin{proof}
\begin{enumerate}
\item Since $x_{i} \leq \delta(x_{i})_{\mathcal{B}}$ for any $i \in I$, $\bigvee_{i \in I} x_{i} \leq \bigvee_{i \in I} \delta(x_{i})_{\mathcal{B}}$. Thus $\delta \left( \bigvee_{i \in I} x_{i} \right)_{\mathcal{B}} \leq \bigvee_{i \in I} \delta(x_{i})_{\mathcal{B}}$.

Since $\delta \left( \bigvee_{i \in I} x_{i} \right)_{\mathcal{B}} \geq \delta(x_{i})_{\mathcal{B}}$ for any $i \in I$, $\delta \left( \bigvee_{i \in I} x_{i} \right)_{\mathcal{B}} \geq \bigvee_{i \in I} \delta(x_{i})_{\mathcal{B}}$. 
Therefore
\begin{equation}
\label{eq-dasein-0.1}
 \delta \left( \bigvee_{i \in I} x_{i} \right)_{\mathcal{B}} = \bigvee_{i \in I} \delta(x_{i})_{\mathcal{B}}. 
 \end{equation}

\item Let $x$ be an element in $\mathcal{L}$. Since $\mathcal{B}$ is atomic, 
\begin{equation}
\label{eq-dasein-49}
\delta(x)_{\mathcal{B}}=\bigvee \{ a \in \mathbb{A}(\mathcal{B}) | a \leq \delta(x)_{\mathcal{B}} \}. 
\end{equation}

By Equations (\ref{eq-dasein-0.1}) and (\ref{eq-dasein-49}),
\begin{equation} 
\label{eq-dasein-50}
\delta \left( \delta(x)_{\mathcal{B}} \right)_{\mathcal{B}'} = \bigvee \{ \delta(a)_{\mathcal{B}'} | a \leq \delta(x)_{\mathcal{B}}, a \in \mathbb{A}(\mathcal{B}) \}. 
\end{equation}
For any $y' \in \mathcal{B}'$ such that $y' \geq x$, $y' \geq \delta(x)_{\mathcal{B}}$ since $\delta(x)_{\mathcal{B}}=\bigwedge \{ y \in \mathcal{B}|y \geq x \}$ and $\mathcal{B}' \subseteq \mathcal{B}$. Thus $y' \geq x$ is equivalent to $y' \geq \delta(x)_{\mathcal{B}}$ for any $y' \in \mathcal{B}'$. Therefore
\begin{equation}
\label{eq-dasein-51}
\begin{split}
\delta \left( \delta(x)_{\mathcal{B}} \right)_{\mathcal{B}'} 
&= \bigwedge \{ y' \in \mathcal{B}' |y' \geq \delta(x)_{\mathcal{B}} \} \\
&= \bigwedge \{ y' \in \mathcal{B}' | y' \geq x \} \\
&= \delta \left( x \right)_{\mathcal{B}'}.
\end{split}
\end{equation}
By Equations (\ref{eq-dasein-50}) and (\ref{eq-dasein-51}),
\begin{equation}
\delta(x)_{\mathcal{B}'} = \bigvee \{\delta(a)_{\mathcal{B}'} | a \leq \delta(x)_{\mathcal{B}}, a \in \mathbb{A}(\mathcal{B}) \}.
\end{equation}

\item Let $a$ be an atom in $\mathcal{B}$, let $y$ be an element in $\mathcal{B}'$ such that 
\begin{equation}
\label{eq-dasein-100}
0 \leq y < \delta(a)_{\mathcal{B}'}, 
\end{equation}
and let 
\begin{equation}
\label{eq-dasein-101}
y' := \delta(a)_{\mathcal{B}'} \wedge y^{\perp}. 
\end{equation}
Then 
\begin{equation}
\label{eq-dasein-102}
\delta(a)_{\mathcal{B}'} = y \vee y'
\end{equation}
 and $0 < y' \in \mathcal{B}'$.

Suppose that $a \wedge y = a \wedge y' =0$. Since $a \leq \delta(a)_{\mathcal{B}'}$,
\begin{equation}
a=a \wedge \delta(a)_{\mathcal{B}'}=a \wedge (y \vee y')=(a \wedge y) \vee (a \wedge y')=0
\end{equation}
 by Equation (\ref{eq-dasein-102}). It is a contradiction. So $a \leq y < \delta(a)_{\mathcal{B}'}$ or $a \leq y' \leq \delta(a)_{\mathcal{B}'}$ since $a$ is an atom in $\mathcal{B}$. $\delta(a)_{\mathcal{B}'}=\bigwedge \{ z \in \mathcal{B}' | a \leq z \}$ and $y, y' \in \mathcal{B}'$ imply $y = \delta(a)_{\mathcal{B}'}$ or $y' = \delta(a)_{\mathcal{B}'}$. By Equation (\ref{eq-dasein-100}), $y'=\delta(a)_{\mathcal{B}'}$. By Equations (\ref{eq-dasein-101}) and (\ref{eq-dasein-102})
\begin{equation}
\label{eq-dasein-103}
\delta(a)_{\mathcal{B}'}=\delta(a)_{\mathcal{B}'} \wedge y^{\perp}, 
\end{equation}
\begin{equation}
\label{eq-dasein-104}
\delta(a)_{\mathcal{B}'}=y \vee \delta(a)_{\mathcal{B}'}.
\end{equation}
Since $\delta(a)_{\mathcal{B}'}^{\perp}=\delta(a)_{\mathcal{B}'}^{\perp} \vee y$ by Equation (\ref{eq-dasein-103}),
\begin{equation}
\begin{split}
y&=y \vee \left( \delta(a)_{\mathcal{B}'} \wedge \delta(a)_{\mathcal{B}'}^{\perp} \right)=\left(y \vee \delta(a)_{\mathcal{B}'} \right) \wedge \left( y \vee \delta(a)_{\mathcal{B}'}^{\perp} \right) \\
&=\delta(a)_{\mathcal{B}'} \wedge \delta(a)_{\mathcal{B}'}^{\perp}=0.
\end{split}
\end{equation}

Therefore $\delta(a)_{\mathcal{B}'}$ is an atom in $\mathcal{B}'$.
\end{enumerate}
\end{proof}

Similar properties that hold in spectral presheaves \cite[Theorem 3 and Theorem 16]{doring2010thing} hold in completely additive spectral presheaves.

\begin{proposition}
\label{prop-daseinisation}

Let $\mathcal{L}$ be a complete orthomodular lattice with the completely additive spectral presheaf $\text{ca}\underline{\Sigma}$, let $x$ be an element in $\mathcal{L}$, let $\mathbb{B}(\mathcal{L})$ be the set of complete atomic Boolean subalgebras in $\mathcal{L}$, and let $\underline{\delta(x)}$ be a mapping from $\mathbb{B}(\mathcal{L})$ to $\textbf{Sets}$ such that
\begin{equation}
 \underline{\delta(x)}(\mathcal{B}) := \alpha_{\mathcal{B}}(\delta(x)_{\mathcal{B}}),
 \end{equation}
where $\alpha_{\mathcal{B}}$ is the mapping defined in Proposition \ref{prop-element}.
Then 
\begin{enumerate}
\item $\underline{\delta(x)}$ is a subobject of $\text{ca}\underline{\Sigma}$.
\item $\underline{\delta(x)}(i_{\mathcal{B}'\mathcal{B}})(\underline{\delta(x)}(\mathcal{B}))=\underline{\delta(x)}(\mathcal{B}')$
\item $\underline{\delta \left( \bigvee_{i \in I} x_{i} \right)} = \bigvee_{i \in I} \underline{\delta(x_{i})}$.
\end{enumerate}
\end{proposition}

\begin{proof}
\begin{enumerate}
\item Let $\mathcal{B}$ and $\mathcal{B}'$ be elements in $\mathbb{B}(\mathcal{L})$ such that $\mathcal{B}' \subseteq \mathcal{B}$.
\begin{equation}
\delta(x)_{\mathcal{B}'} = \bigwedge \{ y \in \mathcal{B}' | y \geq x \} \geq \bigwedge \{ y \in \mathcal{B} | y \geq x \} = \delta(x)_{\mathcal{B}}. 
\end{equation}
For any $\lambda \in \alpha_{\mathcal{B}}(\delta(x)_{\mathcal{B}})$,
\begin{equation} 
\label{eq-dasein-0.001}
\lambda |_{\mathcal{B}'}(\delta(x)_{\mathcal{B}'}) = \lambda(\delta(x)_{\mathcal{B}'}) \geq \lambda(\delta(x)_{\mathcal{B}}) = 1. 
\end{equation}
It shows that
\begin{equation}
\label{eq-dasein-0.01}
\begin{split}
 \underline{\delta(x)}(i_{\mathcal{B}'\mathcal{B}})(\underline{\delta(x)}(\mathcal{B}))
&=\underline{\Sigma}(i_{\mathcal{B}'\mathcal{B}})(\alpha_{\mathcal{B}}(\delta(x)_{\mathcal{B}})) \\
&= \{ \lambda |_{\mathcal{B}'} | \lambda \in \alpha_{\mathcal{B}}(\delta(x)_{\mathcal{B}} \} \\
 &\subseteq \alpha_{\mathcal{B}'}(\delta(x)_{\mathcal{B}'})
 =\underline{\delta(x)}(\mathcal{B}')
\end{split}
\end{equation}
Therefore $\underline{\delta(x)}$ is a subobject of $\text{ca}\underline{\Sigma}$.

\item We will show $\underline{\delta(x)}(i_{\mathcal{B}'\mathcal{B}})(\alpha_{\mathcal{B}}(\delta(x)_{\mathcal{B}})) \supseteq \alpha_{\mathcal{B}'}(\delta(x)_{\mathcal{B}'})$.
Let $\lambda' \in \alpha_{\mathcal{B}'}(\delta(x)_{\mathcal{B}'})$, and let $\mathbb{A}(\mathcal{B})$ be the set of atoms in $\mathcal{B}$. By Lemma \ref{lemma-atom}, 
\begin{equation}
\label{eq-dasein-0.011}
\delta(x)_{\mathcal{B}'}=\bigvee \{ \delta(a)_{\mathcal{B}'} | a \leq \delta(x)_{\mathcal{B}}, a \in \mathbb{A}(\mathcal{B}) \}.
\end{equation}
By Equation (\ref{eq-dasein-0.011}),
\begin{equation}
\begin{split}
 1 &= \lambda'(\delta(x)_{\mathcal{B}'})=\lambda' \left( \bigvee \{ \delta(a)_{\mathcal{B}'} | a \leq \delta(x)_{\mathcal{B}}, a \in \mathbb{A}(\mathcal{B}) \} \right) \\
 &= \bigvee \{ \lambda'(\delta(a)_{\mathcal{B}'}) | a \leq \delta(x)_{\mathcal{B}}, a \in \mathbb{A}(\mathcal{B}) \}.
 \end{split}
 \end{equation}
Thus there is an atom $a \in \mathbb{A}(\mathcal{B})$ such that 
\begin{equation}
\label{eq-dasein-1.998}
\lambda'(\delta(a)_{\mathcal{B}'})=1
\end{equation}
 and 
 \begin{equation}
 \label{eq-dasein-1.999}
 a \leq \delta(x)_{\mathcal{B}}.
 \end{equation}
 Since $\delta(a)_{\mathcal{B}'}$ is an atom in $\mathcal{B}'$ by Lemma \ref{lemma-atom}, $\delta(a)_{\mathcal{B}'} \not\leq y'$ is equivalent to $\delta(a)_{\mathcal{B}'} \wedge y' = 0$ for any $y' \in \mathcal{B}'$. By Equation (\ref{eq-dasein-1.998}),
 \begin{equation}
 \label{eq-dasein-1.9991}
0 = \lambda'(\delta(a)_{\mathcal{B}'} \wedge y')=\lambda'(\delta(a)_{\mathcal{B}'}) \wedge \lambda'(y')=\lambda'(y')
 \end{equation}
for any $y' \in \mathcal{B}'$ such that $\delta(a)_{\mathcal{B}'} \wedge y' = 0$.
By Equations (\ref{eq-dasein-1.998}) and (\ref{eq-dasein-1.9991})
\begin{equation}
\lambda'(y') =
\begin{cases}
1 & (\delta(a)_{\mathcal{B}'} \leq y') \\
0 & (\delta(a)_{\mathcal{B}'} \wedge y' = 0)
\end{cases}
\end{equation}
for any $y' \in \mathcal{B}'$.

Define
\begin{equation}
\lambda(y) =
\begin{cases}
1 & (a \leq y) \\
0 & (a \wedge y = 0).
\end{cases}
\end{equation}
for any $y \in \mathcal{B}$. $\lambda$ is a completely additive two-valued homomorphism of $\mathcal{B}$ by Proposition \ref{prop-atom}, and
\begin{equation}
\lambda |_{\mathcal{B}'}=\lambda'. 
\end{equation}
By Equation (\ref{eq-dasein-1.999}),
\begin{equation}
1=\lambda(a) \leq \lambda(\delta(x)_{\mathcal{B}}).
\end{equation}
Thus $\lambda \in \alpha_{\mathcal{B}}(\delta(x)_{\mathcal{B}})$.
Therefore
\begin{equation}
\label{eq-dasein-2}
  \underline{\delta(x)}(i_{\mathcal{B}'\mathcal{B}})(\alpha_{\mathcal{B}}(\delta(x)_{\mathcal{B}})) \supseteq \alpha_{\mathcal{B}'}(\delta(x)_{\mathcal{B}'}) . 
\end{equation}
By Equations (\ref{eq-dasein-0.01}) and (\ref{eq-dasein-2})
\begin{equation}
\label{eq-dasein-2.01}
  \underline{\delta(x)}(i_{\mathcal{B}'\mathcal{B}})(\alpha_{\mathcal{B}}(\delta(x)_{\mathcal{B}})) = \alpha_{\mathcal{B}'}(\delta(x)_{\mathcal{B}'}) . 
\end{equation}
Therefore
\begin{equation}
\underline{\delta(x)}(i_{\mathcal{B}'\mathcal{B}})(\underline{\delta(x)}(\mathcal{B})) = \underline{\delta(x)}(\mathcal{B}')
\end{equation}
\item For any $\mathcal{B} \in \mathbb{B}(\mathcal{L})$,
\begin{equation}
\begin{split}
\underline{\delta \left( \bigvee_{i \in I} x_{i} \right)}(\mathcal{B})&=  \alpha_{\mathcal{B}} \left( \delta \left( \bigvee_{i \in I}x_{i} \right) \right)  \\
&=\alpha_{\mathcal{B}} \left( \bigvee_{i \in I} \delta \left( x_{i} \right) \right)  \ \ \ (\because \text{Lemma \ref{lemma-atom}} ) \\
&=\bigcup_{i \in I}\alpha_{\mathcal{B}} \left(  \delta \left( x_{i} \right) \right)  \ \ \ (\because \text{Proposition \ref{prop-element}} ) \\
&=\bigcup_{i \in I} \underline{\delta(x_{i})}(\mathcal{B}) \\
&=\left( \bigvee_{i \in I} \underline{\delta(x_{i})} \right) (\mathcal{B}).
\end{split} 
\end{equation}
Therefore
\begin{equation}
\underline{\delta \left( \bigvee_{i \in I} x_{i} \right)} = \bigvee_{i \in I} \underline{\delta(x_{i})}. 
\end{equation}

\end{enumerate}
\end{proof}

Let $\underline{S} \in \text{Subca}\underline{\Sigma}$. $\underline{S}(i_{\mathcal{B}'\mathcal{B}})$ is not necessarily surjective. Condition 2 in Proposition \ref{prop-daseinisation} shows that $\underline{\delta(x)}(i_{\mathcal{B}'\mathcal{B}})$ is always surjective for any element $x \in \mathcal{L}$.

The following fact about negations holds.

\begin{lemma}
\label{lemma-negations-1}
\cite[Lemma 1]{doring2016topos}

Let $\mathcal{L}$ be a complete orthomodular lattice with the completely additive spectral presheaf $\text{ca}\underline{\Sigma}$. Then
\begin{equation}
\neg \underline{\delta(x)} \leq \sim \underline{\delta(x)} \leq \underline{\delta(x^{\perp})}.
\end{equation}
\end{lemma}

\begin{proof}
Let $\mathbb{B}(\mathcal{L})$ be the set of complete atomic Boolean subalgebras of $\mathcal{L}$.
For any $\mathcal{B} \in \mathbb{B}(\mathcal{L})$, it holds $\neg \underline{\delta(x)}(\mathcal{B}) \subseteq \text{ca}\underline{\Sigma}(\mathcal{B}) \setminus \underline{\delta(x)}(\mathcal{B})$ since $\neg \underline{\delta(x)}(\mathcal{B}) \cap \underline{\delta(x)}(\mathcal{B}) = \emptyset$ by Equation (\ref{eq-heyting-negation-0.01}), while  $\sim \underline{\delta(x)}(\mathcal{B}) \supseteq \text{ca}\underline{\Sigma}(\mathcal{B}) \setminus \underline{\delta(x)}(\mathcal{B})$ since $\sim \underline{\delta(x)}(\mathcal{B}) \cup \underline{\delta(x)}(\mathcal{B}) = \text{ca}\underline{\Sigma}$ by Equation (\ref{eq-co-negation-prop}).

Since $\underline{\delta(x)} \vee \underline{\delta(x^{\perp})}=\underline{\delta(x \vee x^{\perp})}=\underline{\delta(1)}=\text{ca}\underline{\Sigma}$ by Proposition \ref{prop-daseinisation}, $\sim \underline{\delta(x)} \leq \underline{\delta(x^{\perp})}$ by Equation (\ref{eq-co-negation-def}).
\end{proof}

In Theorem \ref{theorem-negations}, we examine the conditions such that $\neg \underline{\delta(x)} =  \underline{\delta(x^{\perp})}$ and $\sim \underline{\delta(x)} = \underline{\delta(x^{\perp})}$.

\begin{definition} \cite[Definition 4.10]{cannon2015generalisation} \cite[p. 162]{doring2016topos} 

Let $\mathcal{L}$ be a complete orthomodular lattice with the completely additive spectral presheaf $\text{ca}\underline{\Sigma}$.
Daseinisation is the map $\underline{\delta}$ from $\mathcal{L}$ to $\text{Subca}\underline{\Sigma}$ such that 
\begin{equation}
\underline{\delta}(x)=\underline{\delta(x)}
\end{equation}
 for any $x \in \mathcal{L}$.
\end{definition}

$\underline{\delta}$ is the mapping from an orthomodular lattice to its completely additive spectral presheaf. Since $\underline{\delta}(x)=\underline{\delta}(x')$ implies
\begin{equation}
x=\bigwedge_{\mathcal{B} \in \mathbb{B}(\mathcal{L})} \underline{\delta(x)}(\mathcal{B})=\bigwedge_{\mathcal{B} \in \mathbb{B}(\mathcal{L})} \underline{\delta(x')}(\mathcal{B})=x',
\end{equation}
$\underline{\delta}$ is injective \cite[Lemma 4.11]{cannon2015generalisation}. Therefore the information of the original orthomodular lattice is not lost by daseinisation.

\section{The upper adjoint of daseinisation}
\label{section-upper-adjoint}

Daseinisation is a mapping from an orthomodular lattice $\mathcal{L}$ to its completely additive spectral presheaf $\text{Subca}\underline{\Sigma}$. Cannon and D\"{o}ring \cite{cannon2015generalisation} pointed out that there is a mapping from $\text{Subca}\underline{\Sigma}$ to $\mathcal{L}$.

For any subset $\{ x_{i} \in \mathcal{L} | i \in I \}$
\begin{equation}
\label{eq-join-preserving-1}
\underline{\delta} \left( \bigvee_{i \in I} x_{i} \right) 
=\underline{\delta \left( \bigvee_{i \in I} x_{i} \right)}
=\bigvee_{i \in I} \underline{ \delta \left(  x_{i} \right)}
=\bigvee_{i \in I} \underline{\delta} \left(  x_{i} \right) 
\end{equation}
by Proposition \ref{prop-daseinisation}. Thus the following map can be defined by the adjoint functor theorem \cite[Theorem 3.3.3 and Example 3.3.9e]{borceux1994handbook1} \cite[Proposition 2.30]{cannon2015generalisation}.

\begin{definition}
\cite[p. 59]{cannon2015generalisation}
Let $\mathcal{L}$ be a complete orthomodular lattice, with the completely additive spectral presheaf $\text{ca}\underline{\Sigma}$. $\varepsilon$ is a map from $\text{Subca} \underline{\Sigma}$ to $\mathcal{L}$ such that
\begin{equation}
 \varepsilon \left( \underline{S} \right) = \bigvee \{ x \in \mathcal{L} | \underline{\delta}(x) \leq \underline{S} \}. 
 \end{equation}
for any $\underline{S} \in \text{Subca} \underline{\Sigma}$. It is called the upper adjoint of $\underline{\delta}$.
\end{definition}

Cannon and D\"{o}ring \cite{cannon2015generalisation} showed the following important results about the upper adjoint of daseinisation. These facts will be used to prove Theorem \ref{theorem-negations}.

\begin{proposition} \cite[Lemma 4.12, Lemma 4.13, and Lemma 4.14]{cannon2015generalisation}
\label{prop-cannon}

Let $\mathcal{L}$ be a complete orthomodular lattice with the completely additive spectral presheaf $\text{ca}\underline{\Sigma}$, and let $\mathbb{B}(\mathcal{L})$ be the set of complete atomic Boolean subalgebras of $\mathcal{L}$. 
\begin{enumerate}
\item The upper adjoint of $\underline{\delta}$ is given by
\begin{equation}
\varepsilon \left( \underline{S} \right) = \bigwedge_{\mathcal{B} \in \mathbb{B}(\mathcal{L})} \alpha^{-1}_{\mathcal{B}} \left( \underline{S} \left( \mathcal{B} \right) \right),
\end{equation}
where $\alpha_{\mathcal{B}}$ is the mapping defined in Proposition \ref{prop-element}.
\item
\begin{equation}
\label{eq-epsilon-100}
\begin{split}
&\varepsilon \circ \underline{\delta} = \text{Id}_{\mathcal{L}}, \\
&\underline{\delta} \circ \varepsilon \leq \text{Id}_{\text{Subca}\underline{\Sigma}}.
\end{split}
\end{equation}
\item For any $\underline{S}$ and $\underline{T}$ in $\text{Subca}\underline{\Sigma}$,
\begin{equation}
\varepsilon(\underline{S}) \wedge \varepsilon(\underline{T}) = \varepsilon(\underline{S} \wedge \underline{T})
\end{equation}

\end{enumerate}
\end{proposition}

\section{Negations and meets}
\label{section-negations-meets}

Equation (\ref{eq-join-preserving-1}) shows that joins are preserved under daseinisation. On the other hand, daseinisation does not necessarily preserve negations and meets.

In Theorem \ref{theorem-negations}, we investigate the conditions under which negations and meets are preserved by daseinisation, and the condition that any element in the Heyting algebra transformed through daseinisation corresponds to an element in the original orthomodular lattice. To prove Theorem \ref{theorem-negations}, we prepare the following lemma.

\begin{lemma}
\label{lemma-neg}
Let $\mathcal{L}$ be an orthomodular lattice with the completely additive spectral presheaf $\text{ca}\underline{\Sigma}$, let $\mathbb{B}(\mathcal{L})$ be the set of complete atomic Boolean subalgebras of $\mathcal{L}$, and let $z$ be an element in $\mathcal{L}$ such that $0 < z < 1$, and let $\underline{\delta}$ be the daseinisation of $\mathcal{L}$ to $\text{Subca}\underline{\Sigma}$.

If there is an element $y \in \mathcal{L} \setminus \{ 0, z, z^{\perp}, 1 \}$, there are an element $x \in \mathcal{L}$ and $\mathcal{B}_{0} \in \mathbb{B}(\mathcal{L})$ such that
\begin{equation}
\neg \underline{\delta(x)}(\mathcal{B}_{0})=\alpha_{\mathcal{B}_{0}}(0), \ \ \ \ \neg \underline{\delta(x)} \neq \underline{\delta(0)}, 
\end{equation}
\begin{equation}
\sim \underline{\delta(x)}(\mathcal{B}_{0})=\alpha_{\mathcal{B}_{0}}(0), \ \ \ \ \sim \underline{\delta(x)} \neq \underline{\delta(0)},
\end{equation}
where $\alpha_{\mathcal{B}_{0}}$ is an isomorphism from $\mathcal{B}_{0}$ to $\text{ca}\underline{\Sigma}(\mathcal{B}_{0})$ which is defined in Proposition \ref{prop-element}.
\end{lemma}

\begin{proof}
Let $y$ be an element in $\mathcal{L}$ such that $y \in \mathcal{L} \setminus \{ 0, z, z^{\perp}, 1 \}$. There are two cases.
\begin{enumerate}[a.]
\item First we examine the case where there are elements $u \in \{y, y^{\perp} \}$ and $v \in \{z, z^{\perp} \}$ such that $u < v$. Let $\mathcal{B}_{v} := \{ 0, v, v^{\perp}, 1 \}$.

Suppose that $u \leq v^{\perp}$. Then $u \leq v \wedge v^{\perp} = 0$. It is a contradiction. Thus $u \not\leq v^{\perp}$. Therefore 
\begin{equation}
\delta(u)_{\mathcal{B}_{v}}=v. 
\end{equation}
Since $v>0$,
\begin{equation}
\label{eq-neg-1}
\underline{\delta((u^{\perp})^{\perp})}(\mathcal{B}_{v}) 
= \underline{\delta(u)}(\mathcal{B}_{v})
=\alpha_{\mathcal{B}_{v}}(\delta(u)_{\mathcal{B}_{v}})
=\alpha_{\mathcal{B}_{v}}(v)
\neq \alpha_{\mathcal{B}_{v}}(0).
\end{equation}

Suppose that $u^{\perp} \leq v$. Then $1 = u \vee u^{\perp} \leq v$. It is a contradiction. Thus $u^{\perp} \not\leq v$. Since $v^{\perp} < u^{\perp}$ and $u^{\perp} \not\leq v$, $\delta(u^{\perp})_{\mathcal{B}_{v}}=1$. Thus
\begin{equation}
\label{eq-neg-2}
\underline{\delta(u^{\perp})}(\mathcal{B}_{v})
=\alpha_{\mathcal{B}_{v}}(\delta(u^{\perp})_{\mathcal{B}_{v}})
=\alpha_{\mathcal{B}_{v}}(1).
\end{equation}
Because the set of complete atomic Boolean  subalgebras of $\mathcal{B}_{v}$ is $\{ \mathcal{B}_{v} \}$,
\begin{equation}
\label{eq-neg-2.1}
\begin{split}
\neg \underline{\delta(u^{\perp})}(\mathcal{B}_{v})
&=\{ \lambda \in \text{ca}\underline{\Sigma}(\mathcal{B}_{v}) | \lambda \not\in \underline{\delta(u^{\perp})}(\mathcal{B}_{v}) \} \\
&=\{ \lambda \in \text{ca}\underline{\Sigma}(\mathcal{B}_{v}) | \lambda \not\in \alpha_{\mathcal{B}_{v}}(1) \} \\
&=\alpha_{\mathcal{B}_{v}}(0).
\end{split}
\end{equation}
By Equation (\ref{eq-co-negation-prop}),
\begin{equation}
\label{eq-neg-2.111}
\begin{split}
\sim \underline{\delta(u^{\perp})}(\mathcal{B}_{v})
&=\{ \lambda \in \text{ca}\underline{\Sigma}(\mathcal{B}_{v}) | \lambda \not\in \underline{\delta(u^{\perp})}(\mathcal{B}_{v}) \} \\
&=\{ \lambda \in \text{ca}\underline{\Sigma}(\mathcal{B}_{v}) | \lambda \not\in \alpha_{\mathcal{B}_{v}}(1) \} \\
&=\alpha_{\mathcal{B}_{v}}(0).
\end{split}
\end{equation}



Let $\mathcal{B}_{u} := \{ 0, u, u^{\perp}, 1 \}$. Then $\neg \underline{\delta(u^{\perp})}(\mathcal{B}_{u}) \neq \alpha_{\mathcal{B}_{u}}(0)$ and $\sim \underline{\delta(u^{\perp})}(\mathcal{B}_{u}) \neq \alpha_{\mathcal{B}_{u}}(0)$. Thus 
\begin{equation}
\neg \underline{\delta(u^{\perp})} \neq \underline{\delta(0)}, \ \ \ \ \sim \underline{\delta(u^{\perp})} \neq \underline{\delta(0)}.
\end{equation}

\item Next we examine the case where $u \not< v$ for any elements $u \in \{y, y^{\perp} \}$ and $v \in \{z, z^{\perp} \}$. Since $\mathcal{L}$ is an orthomodular lattice,
\begin{equation}
\label{eq-wedge-1}
u=(u \wedge v) \vee (u \wedge (u \wedge v)^{\perp}).
\end{equation}
If $u \wedge (u \wedge v)^{\perp}=0$, then $u = u \wedge v \leq v$ by Equation (\ref{eq-wedge-1}). It is a contradiction since $u \not< v$. Thus $u \wedge (u \wedge v)^{\perp} \neq 0$, which implies $(u \wedge v)^{\perp} > 0$.

Thus
\begin{equation}
 (y \wedge z)^{\perp} \wedge (y \wedge z^{\perp})^{\perp} \wedge (y^{\perp} \wedge z)^{\perp} \wedge (y^{\perp} \wedge z^{\perp})^{\perp} > 0. 
\end{equation}
Therefore
\begin{equation}
\label{eq-wedge-2}
(y \wedge z) \vee (y \wedge z^{\perp}) \vee (y^{\perp} \wedge z) \vee (y^{\perp} \wedge z^{\perp}) < 1. 
\end{equation}
By Equation (\ref{eq-wedge-2}) and Proposition \ref{prop-commute}, $y$ does not commute with $z$.

Let $\mathcal{B}_{z} := \{ 0, z, z^{\perp}, 1 \}$. Then 
\begin{equation}
\label{eq-wedge-3}
\delta(y)_{\mathcal{B}_{z}}=1, \ \ \ \ \  \delta(y^{\perp})_{\mathcal{B}_{z}}=1.
\end{equation}
Therefore
\begin{equation}
\label{eq-wedge-4.1}
\begin{split}
&\underline{\delta(y)}(\mathcal{B}_{z})=\alpha_{\mathcal{B}_{z}}(\delta(y)_{\mathcal{B}_{z}})=\alpha_{\mathcal{B}_{z}}(1), \\
&\underline{\delta(y^{\perp})}(\mathcal{B}_{z})=\alpha_{\mathcal{B}_{z}}(\delta(y^{\perp})_{\mathcal{B}_{z}})=\alpha_{\mathcal{B}_{z}}(1).
\end{split}
\end{equation}
Because the set of complete atomic Boolean  subalgebras of $\mathcal{B}_{z}$ is $\{ \mathcal{B}_{z} \}$,
\begin{equation}
\label{eq-wedge-4}
\begin{split}
\neg \underline{\delta(y)}(\mathcal{B}_{z})
&=\{ \lambda \in \text{ca}\underline{\Sigma}(\mathcal{B}_{z}) | \lambda \not\in \underline{\delta(y)}(\mathcal{B}_{z}) \} \\
&=\{ \lambda \in \text{ca}\underline{\Sigma}(\mathcal{B}_{z}) | \lambda \not\in \alpha_{\mathcal{B}_{z}}(1) \} \\
&=\alpha_{\mathcal{B}_{z}}(0).
\end{split}
\end{equation}

By Equation (\ref{eq-co-negation-prop}),
\begin{equation}
\label{eq-neg-2.11}
\begin{split}
\sim \underline{\delta(y)}(\mathcal{B}_{z})
&=\{ \lambda \in \text{ca}\underline{\Sigma}(\mathcal{B}_{z}) | \lambda \not\in \underline{\delta(y)}(\mathcal{B}_{z}) \} \\
&=\{ \lambda \in \text{ca}\underline{\Sigma}(\mathcal{B}_{z}) | \lambda \not\in \alpha_{\mathcal{B}_{z}}(1) \} \\
&=\alpha_{\mathcal{B}_{z}}(0).
\end{split}
\end{equation}


Let $\mathcal{B}_{y} := \{ 0, y, y^{\perp}, 1 \}$. Then $\neg \underline{\delta(y)}(\mathcal{B}_{y}) \neq \alpha_{\mathcal{B}_{y}}(0)$ and $\sim \underline{\delta(y)}(\mathcal{B}_{y}) \neq \alpha_{\mathcal{B}_{y}}(0)$. Thus 
\begin{equation}
\neg \underline{\delta(y)} \neq \underline{\delta(0)}, \ \ \ \ \sim \underline{\delta(y)} \neq \underline{\delta(0)}.
\end{equation}
\end{enumerate}
\end{proof}

\begin{theorem}
\label{theorem-negations}
Let $\mathcal{L}$ be an orthomodular lattice with the completely additive spectral presheaf $\text{ca}\underline{\Sigma}$, and let $z$ be an element in $\mathcal{L}$ such that $0 < z < 1$, let $\underline{\delta}$ be the daseinisation of $\mathcal{L}$ to $\text{Subca}\underline{\Sigma}$, and let $\varepsilon$ be the upper adjoint of $\underline{\delta}$.

 The following conditions are equivalent:

\begin{enumerate}
\item For any $x \in \mathcal{L}$, $\neg \underline{\delta}(x) = \underline{\delta}(x^{\perp})$,
\item For any $x \in \mathcal{L}$, $\sim \underline{\delta}(x) = \underline{\delta}(x^{\perp})$,
\item For any $x, y \in \mathcal{L}$, $\underline{\delta}(x) \wedge \underline{\delta}(y) = \underline{\delta}(x \wedge y)$,
\item For any $x \in \mathcal{L}$, there is an element $u$ in $\mathcal{L}$ such that $\neg \underline{\delta}(x) = \underline{\delta}(u)$,
\item For any $x \in \mathcal{L}$, there is an element $u$ in $\mathcal{L}$ such that $\sim \underline{\delta}(x) = \underline{\delta}(u)$,
\item For any $x, y \in \mathcal{L}$, there is an element $u$ in $\mathcal{L}$ such that $\underline{\delta}(x) \wedge \underline{\delta}(y) = \underline{\delta}(u)$,
\item $\underline{\delta} \circ \varepsilon = \text{Id}_{\text{Subca}\underline{\Sigma}}$,
\item $\mathcal{L}=\{ 0, z, z^{\perp}, 1 \}$.
\end{enumerate}
\end{theorem}

\begin{proof}
Let $\mathbb{B}(\mathcal{L})$ be the set of complete atomic Boolean subalgebras of $\mathcal{L}$ and let $\alpha_{\mathcal{B}}$ be the mapping defined in Proposition \ref{prop-element}.

\begin{enumerate}[I.]
\item

\begin{description}
\item[$1 \Longrightarrow 4$]
Trivial.
\item[$4 \Longrightarrow 8$]
Suppose that there is an element $y$ in $\mathcal{L}$ such that $y \in \mathcal{L} \setminus \{ 0, z, z^{\perp}, 1 \}$ under Condition 4. By Lemma \ref{lemma-neg}, there are an element $x \in \mathcal{L}$ and $\mathcal{B}_{0} \in \mathbb{B}(\mathcal{L})$ such that
\begin{equation}
\label{eq-neg-existence-1}
\neg \underline{\delta(x)}(\mathcal{B}_{0})=\alpha_{\mathcal{B}_{0}}(0), \ \ \ \ \neg \underline{\delta(x)} \neq \underline{\delta(0)}.
\end{equation}
By Condition 4, there is an element $u \in \mathcal{L}$ such that 
\begin{equation}
\label{eq-neg-existence-1.01}
\neg \underline{\delta(x)}=\underline{\delta(u)}. 
\end{equation}
By Equations (\ref{eq-neg-existence-1}) and (\ref{eq-neg-existence-1.01}), 
\begin{equation}
\underline{\delta(u)} \neq \underline{\delta(0)}. 
\end{equation}
Since $\underline{\delta}$ is injective,
\begin{equation}
u \neq 0. 
\end{equation}
On the other hand,
\begin{equation}
\begin{split}
u&=\varepsilon \circ \underline{\delta}(u)
=\varepsilon \circ \underline{\delta(u)}
=\varepsilon \circ \neg \underline{\delta(x)} \\
&=\bigwedge_{\mathcal{B} \in \mathbb{B}(\mathcal{L})} \alpha_{\mathcal{B}}^{-1}(\neg \underline{\delta(x)}(\mathcal{B})) \\
&\leq \alpha_{\mathcal{B}_{0}}^{-1}(\neg \underline{\delta(x)}(\mathcal{B}_{0})) \\
&=\alpha_{\mathcal{B}_{0}}^{-1}(\alpha_{\mathcal{B}_{0}}(0))
=0
\end{split}
\end{equation}
by Proposition \ref{prop-cannon} and Equation (\ref{eq-neg-existence-1}). It is a contradiction. Therefore $\mathcal{L} = \{ 0, z, z^{\perp}, 1 \}$.

\item[$2 \Longrightarrow 5$]
Trivial.
\item[$5 \Longrightarrow 8$]
It can be shown similarly to the proof of $4 \Longrightarrow 8$.

\item[$6 \Longrightarrow 3$] \cite[p. 1174]{eva2017topos}

Let $x$, $y$, and $u$ be elements in $\mathcal{L}$ such that $\mathcal{L}$ such that $\underline{\delta(x)} \wedge \underline{\delta(y)} = \underline{\delta(u)}$. By Proposition \ref{prop-cannon},
\begin{equation}
\begin{split}
u
&=\varepsilon \circ \underline{\delta}(u)
=\varepsilon \circ \underline{\delta(u)}
=\varepsilon(\underline{\delta(x)} \wedge \underline{\delta(y)}) \\
&=\varepsilon(\underline{\delta(x)}) \wedge \varepsilon(\underline{\delta(y)})=x \wedge y.
\end{split}
\end{equation}
Thus $\underline{\delta(x)} \wedge \underline{\delta(y)} = \underline{\delta(x \wedge y)}$.

\item[$3 \Longrightarrow 8$]
Let $y$ be an element in $\mathcal{L}$ such that $y \in \mathcal{L} \setminus \{ 0, z, z^{\perp}, 1 \}$.

There are two cases.
\begin{enumerate}[a.]
\item First we examine the case where there are elements $u \in \{ y, y^{\perp} \}$ and $v \in \{ z, z^{\perp} \}$ such that $u \vee v  < 1$. 

Since $\mathcal{L}$ is an orthomodular lattice,
\begin{equation}
u \vee v = u \vee (u^{\perp} \wedge (u \vee v)).
\end{equation}
Let $v' := u^{\perp} \wedge (u \wedge v)$. Then 
\begin{equation}
\label{eq-wedge-5}
u \vee v' = u \vee v < 1 , \ \ \ \ \ u \wedge v' = 0.
\end{equation}

Let $\mathcal{B}_{u \vee v'}=\{ 0, u \vee v', (u \vee v')^{\perp}, 1 \}$. Since $\delta(u)_{\mathcal{B}_{u \vee v'}}=u \vee v'$ and $\delta(v')_{\mathcal{B}_{u \vee v'}}=u \vee v'$,
\begin{equation}
\label{eq-wedge-6}
\begin{split}
&\underline{\delta(u)}(\mathcal{B}_{u \vee v'})=\alpha_{\mathcal{B}_{u \vee v'}}(\delta(u)_{\mathcal{B}_{u \vee v'}})=\alpha_{\mathcal{B}_{u \vee v'}}(u \vee v'), \\
&\underline{\delta(v')}(\mathcal{B}_{u \vee v'})=\alpha_{\mathcal{B}_{u \vee v'}}(\delta(v')_{\mathcal{B}_{u \vee v'}})=\alpha_{\mathcal{B}_{u \vee v'}}(u \vee v').
\end{split}
\end{equation}
By Equation (\ref{eq-wedge-6}),
\begin{equation}
\label{eq-wedge-7}
\begin{split}
\underline{\delta(u)}(\mathcal{B}_{u \vee v'}) \cap \underline{\delta(v')}(\mathcal{B}_{u \vee v'})
&=\alpha_{\mathcal{B}_{u \vee v'}}(u \vee v')  \\
&\neq \alpha_{\mathcal{B}_{u \vee v'}}(0)
\end{split}
\end{equation}
because $0 < u \leq u \vee v'$.

By Equation (\ref{eq-wedge-5}),
\begin{equation}
\label{eq-wedge-8}
\begin{split}
\underline{\delta(u \wedge v')}(\mathcal{B}_{u \vee v'}) 
&=\alpha_{\mathcal{B}_{u \vee v'}}(\delta(u \wedge v')_{\mathcal{B}_{u \vee v'}}) \\
&=\alpha_{\mathcal{B}_{u \vee v'}}(\delta(0)_{\mathcal{B}_{u \vee v'}}) \\
&=\alpha_{\mathcal{B}_{u \vee v'}}(0).
\end{split}
\end{equation}
Therefore
\begin{equation}
\label{eq-wedge-9}
\underline{\delta(u)} \wedge \underline{\delta(v')} \neq \underline{\delta(u \wedge v')}.
\end{equation}
\item
Next we examine the case where $u \vee v = 1$ for any $u \in \{ y, y^{\perp} \}$ and $v \in \{ z, z^{\perp} \}$.

By this condition, $u^{\perp} \vee v^{\perp}=1$, that is, 
\begin{equation}
\label{eq-wedge-10}
u \wedge v = 0
\end{equation}
 for any $u \in \{ y, y^{\perp} \}$ and $v \in \{ z, z^{\perp} \}$.

Since $u^{\perp} \vee v=1$,
\begin{equation}
\label{eq-wedge-11}
u \wedge (u^{\perp} \vee v)=u \wedge 1 = u > 0
\end{equation}
 for any $u \in \{ y, y^{\perp} \}$ and $v \in \{ z, z^{\perp} \}$.
By Equations (\ref{eq-wedge-10}), (\ref{eq-wedge-11}), and Proposition \ref{prop-commute}, $u$ does not commute with $v$  for any $u \in \{ y, y^{\perp} \}$ and $v \in \{ z, z^{\perp} \}$.

Let $\mathcal{B}_{z} := \{ 0, z, z^{\perp}, 1 \}$. Since $\delta(y)_{\mathcal{B}_{z}}=\delta(y^{\perp})_{\mathcal{B}_{z}}=1$,
\begin{equation}
\label{eq-wedge-12}
\begin{split}
&\underline{\delta(y)}(\mathcal{B}_{z})=\alpha_{\mathcal{B}_{z}}(\delta(y)_{\mathcal{B}_{z}})=\alpha_{\mathcal{B}_{z}}(1), \\
&\underline{\delta(y^{\perp})}(\mathcal{B}_{z})=\alpha_{\mathcal{B}_{z}}(\delta(y^{\perp})_{\mathcal{B}_{z}})=\alpha_{\mathcal{B}_{z}}(1).
\end{split}
\end{equation}
Thus
\begin{equation}
\label{eq-wedge-13}
\underline{\delta(y)}(\mathcal{B}_{z}) \cap \underline{\delta(y^{\perp})}(\mathcal{B}_{z})=\alpha_{\mathcal{B}_{z}}(1).
\end{equation}
On the other hand,
\begin{equation}
\label{eq-wedge-14}
\begin{split}
\underline{\delta(y \wedge y^{\perp})}(\mathcal{B}_{z})
&=\alpha_{\mathcal{B}_{z}}(\delta(y \wedge y^{\perp}))_{\mathcal{B}_{z}}) \\
&=\alpha_{\mathcal{B}_{z}}(\delta(0))_{\mathcal{B}_{z}}) \\
&=\alpha_{\mathcal{B}_{z}}(0).
\end{split}
\end{equation}
By Equations (\ref{eq-wedge-13}) and (\ref{eq-wedge-14}),
\begin{equation}
\underline{\delta(y)} \wedge \underline{\delta(y^{\perp})} \neq \underline{\delta(y \wedge y^{\perp})}.
\end{equation}
\end{enumerate}

\item[$7 \Longrightarrow 8$]
By Lemma \ref{lemma-neg}, there are an element $x \in \mathcal{L}$ and $\mathcal{B}_{0} \in \mathbb{B}_{0}$ such that
\begin{equation}
\label{eq-epsilon-0.99}
\neg \underline{\delta(x)}(\mathcal{B}_{0}) = \alpha_{\mathcal{B}_{0}}(0), \ \ \ \ \ \neg \underline{\delta(x)} \neq \underline{\delta(0)}.
\end{equation}
By Proposition \ref{prop-cannon} and Equation (\ref{eq-epsilon-0.99}),
\begin{equation}
\label{eq-epsilon-1}
\begin{split}
\varepsilon(\neg \underline{\delta(x)})
&=\bigwedge_{\mathcal{B} \in \mathbb{B}(\mathcal{L})} \alpha_{\mathcal{B}}^{-1}(\neg \underline{\delta(x)}(\mathcal{B})) \\
&\leq \alpha_{\mathcal{B}_{0}}^{-1}(\neg \underline{\delta(x)}(\mathcal{B}_{0})) = \alpha_{\mathcal{B}_{0}}^{-1}(\alpha_{\mathcal{B}_{0}}(0))=0.
\end{split}
\end{equation}
By Equations (\ref{eq-epsilon-1}) and (\ref{eq-epsilon-0.99}),
\begin{equation}
\underline{\delta} \circ \varepsilon(\neg \underline{\delta(x)}) = \underline{\delta(0)} \neq \neg \underline{\delta(x)}.
\end{equation}
By Equation (\ref{eq-epsilon-100}),
\begin{equation}
\underline{\delta} \circ \varepsilon < \text{Id}_{\text{Subca}\underline{\Sigma}}.
\end{equation}
\end{description}
\item Let $\mathcal{L} = \{ 0, z, z^{\perp}, 1 \}$. Then $\mathbb{B}(\mathcal{L})=\{ \mathcal{L} \}$.
\begin{description}
\item[$8 \Longrightarrow 1$]
For any $x \in \{0, z, z^{\perp}, 1 \}$,
\begin{equation}
\begin{split}
\neg \underline{\delta(x)}(\mathcal{L})
&=\{ \lambda \in \text{ca}\underline{\Sigma}(\mathcal{L}) | \lambda \not\in \underline{\delta(x)}(\mathcal{L}) \} \\
&=\{ \lambda \in \text{ca}\underline{\Sigma}(\mathcal{L}) | \lambda \not\in \alpha_{\mathcal{L}}(x) \} \\
&=\alpha_{\mathcal{L}}(x^{\perp}) \\
&=\underline{\delta(x^{\perp})}(\mathcal{L})
\end{split}
\end{equation}
implies
\begin{equation}
\label{eq-epsilon-3}
\neg \underline{\delta(x)}=\underline{\delta(x^{\perp})}, 
\end{equation}
\item[$1 \Longrightarrow 2$] It holds by Lemma \ref{lemma-negations-1}.
\item[$8 \Longrightarrow 3$]
\begin{equation}
\begin{split}
\underline{\delta(z)}(\mathcal{L}) \cap \underline{\delta(z^{\perp})}(\mathcal{L}) 
&= \alpha_{\mathcal{L}}(z) \cap \alpha_{\mathcal{L}}(z^{\perp})
=\alpha_{\mathcal{L}}(0) \\
&=\underline{\delta(0)}(\mathcal{L}) 
=\underline{\delta(z \wedge z^{\perp})}(\mathcal{L})
\end{split}
\end{equation}
implies
\begin{equation}
\label{eq-epsilon-4}
\underline{\delta(z)} \wedge \underline{\delta(z^{\perp})} = \underline{\delta(z \wedge z^{\perp})}.
\end{equation}
\item[$3 \Longrightarrow 6$] Trivial.
\item[$8 \Longrightarrow 7$]
By Equations (\ref{eq-epsilon-3}) and (\ref{eq-epsilon-4}), the set of all subobjects is 
\begin{equation}
\{ \underline{\delta(0)}, \underline{\delta(z)}, \underline{\delta(z^{\perp})}, \underline{\delta(1)} \}. 
\end{equation}
By Proposition \ref{prop-cannon}
\begin{equation}
\label{eq-epsilon-5}
\varepsilon \circ \underline{\delta(x)}
=\varepsilon \circ \underline{\delta}(x)
= x
\end{equation}
for any element $x$ in $\{ 0, z, z^{\perp}, 1 \}$.
By Equation (\ref{eq-epsilon-5})
 \begin{equation}
 \underline{\delta} \circ \varepsilon \circ \underline{\delta(x)}= \underline{\delta}(x)= \underline{\delta(x)}.
 \end{equation}
 Therefore
 \begin{equation}
\underline{\delta} \circ  \varepsilon= \text{Id}_{\text{Subca}\underline{\Sigma}}.
 \end{equation}
\end{description}
\end{enumerate}
\end{proof}

Cannon and D\"{o}ring \cite{cannon2015generalisation} used complete Boolean algebras to construct presheaves. Theorem \ref{theorem-negations} holds in this case as well.


Let $\mathcal{L}$ be a non-distributive orthomodular lattice, or a complete Boolean algebra with more than four elements. According to Theorem \ref{theorem-negations}, there is an element $x$ in $\mathcal{L}$ such that for any element $u$ in $\mathcal{L}$ such that $\neg \underline{\delta}(x) \neq \underline{\delta}(u)$. Similarly there are elements $x$ and $y$ in $\mathcal{L}$ such that for any element $u$ in $\mathcal{L}$ such that $\underline{\delta}(x) \wedge \underline{\delta}(y) \neq \underline{\delta}(u)$. Therefore there is an element in the Heyting algebra transformed through daseinisation that does not correspond to any element in the original orthomodular lattice $\mathcal{L}$. 

\section{Concluding Remarks}
\label{section-concluding}
Based on the properties of the complete atomic Boolean algebras examined in Section \ref{section-complete-atomic-boolean}, the properties of daseinisation, which is a mapping from an orthomodular lattice to its completely additive spectral presheaf, are investigated in Section \ref{section-daseinisation}. Daseinisation in the present paper has the same properties as daseinisation defined by D\"{o}ring and Isham \cite{doring2008topos2,doring2010thing,cannon2015generalisation} in the case of spectral presheaf (Proposition \ref{prop-daseinisation}).

In Section \ref{section-negations-meets}, we consider two problems raised in Section \ref{section-introduction}. One of the problems is under what conditions daseinisation does not preserve negations and meets, and the other is under what conditions the Heyting algebra transformed from the orthomodular lattice by daseinisation contains an element that does not correspond to the element of the original orthomodular lattice. In  Theorem \ref{theorem-negations}, we examine these two questions. The results of Theorem \ref{theorem-negations} hold not only for completely additive spectral presheaves but also for spectral presheaves.

The answer to these problems is that not only in non-distributive orthomodular lattices but also in Boolean algebras with more than four elements, daseinisation does not preserve negations and meets, and the Heyting algebra transformed from an orthomodular lattice by daseinisation contains an element that does not correspond to any element of the original orthomodular lattice.
Furthermore, it is shown that the conditions under which negations and meets are preserved by daseinisation are equivalent to the condition that any element in the Heyting algebra transformed through daseinisation corresponds to an element in the original orthomodular lattice. This shows that the two issues are closely related. The answer to the second question suggests that the fact that daseinisation introduces an element that does not correspond to any element of the original orthomodular lattice is not due to a quantum property but to a property of daseinisation itself because this fact occurs even in the case of a Boolean algebra containing more than four elements.

\bibliographystyle{plain}

\bibliography{kitajima}
%
%

\end{document}